# GFSR-Net: Guided Focus via Segment-Wise Relevance Network for Interpretable Deep Learning in Medical Imaging

Jhonatan Contreras1,2 and Thomas Bocklitz1,2,*

1   Institute of Physical Chemistry (IPC) and Abbe Center of Photonics (ACP), Friedrich Schiller University Jena, Member of the Leibniz Centre for Photonics in Infection Research (LPI), Helmholtzweg 4, 07743 Jena, Germany.

2   Leibniz Institute of Photonic Technology, Member of Leibniz Health Technologies, Member of the Leibniz. Centre for Photonics in Infection Research (LPI), Albert Einstein Straße 9, 07745 Jena, Germany.

*   Correspondence: Thomas.bocklitz@uni-jena.de

## Abstract

Deep learning has achieved remarkable success in medical image analysis, however its adoption in clinical practice is limited by a lack of interpretability. These models often make correct predictions without explaining their reasoning. They may also rely on image regions unrelated to the disease or visual cues, such as annotations, that are not present in real-world conditions. This can reduce trust and increase the risk of misleading diagnoses. We introduce the Guided Focus via Segment-Wise Relevance Network (GFSR-Net), an approach designed to improve interpretability and reliability in medical imaging.  GFSR-Net uses a small number of human annotations to approximate where a person would focus within an image intuitively, without requiring precise boundaries or exhaustive markings, making the process fast and practical. During training, the model learns to align its focus with these areas, progressively emphasizing features that carry diagnostic meaning. This guidance works across different types of natural and medical images, including chest X-rays, retinal scans, and dermatological images. Our experiments demonstrate that GFSR achieves comparable or superior accuracy while producing saliency maps that better reflect human expectations. This reduces the reliance on irrelevant patterns and increases confidence in automated diagnostic tools.

## 1. Introduction

Deep learning has achieved remarkable results in medical imaging tasks, including disease classification, lesion detection, and anatomical segmentation [1], [2]. However, despite these advancements, its use in clinical settings remains limited. One significant barrier is the lack of interpretability. In healthcare, where diagnostic decisions can have serious consequences, models must be accurate and provide clear, medically sound explanations for their predictions. This necessity has led to a growing interest in explainable artificial intelligence (XAI), which seeks to clarify the reasoning behind a model's predictions and build confidence in its outcomes [3], [4].

One of the most widely used XAI methods is Gradient-weighted Class Activation Mapping (Grad-CAM) [5], which generates heatmaps by computing the gradient of the output with respect to convolutional features. These heatmaps highlight the regions of an image that most influence a model's prediction. However, like most XAI methods, Grad-CAM only provides a post hoc explanation. That is, it operates after training and does not affect the learning process. Consequently, while Grad-CAM can reveal when a model relies on irrelevant features, it cannot prevent this behavior during parameters optimization.

To illustrate this limitation in practice, we trained convolutional neural networks (CNNs) on chest X-ray [6] images for pneumonia classification, RetinaMNIST [7] for diabetic retinopathy grading, and DermaMNIST-C [8] for dermatological disease classification.

Grad-CAM visualizations reveal that the models sometimes rely on features unrelated to the underlying pathology across all three datasets. In chest X-rays, for example, the attention maps sometimes highlight non-diagnostic regions, such as the neck, head, image borders, or the "R" marker. This introduces bias and privileged information that is absent in real testing conditions (Figure 1a). In retinal images, the model frequently focuses on the optic disc or corner annotations (Figure 1b). These structures are not indicative of disease severity and reduce diagnostic reliability. For dermatological images, Grad-CAM highlights often extend to background textures, hair, or skin outside the lesion area rather than the lesion itself (Figure 1c). These findings suggest that standard CNNs may rely on irrelevant visual cues rather than clinically meaningful features, which undermines the robustness and trustworthiness of automated diagnostic systems.

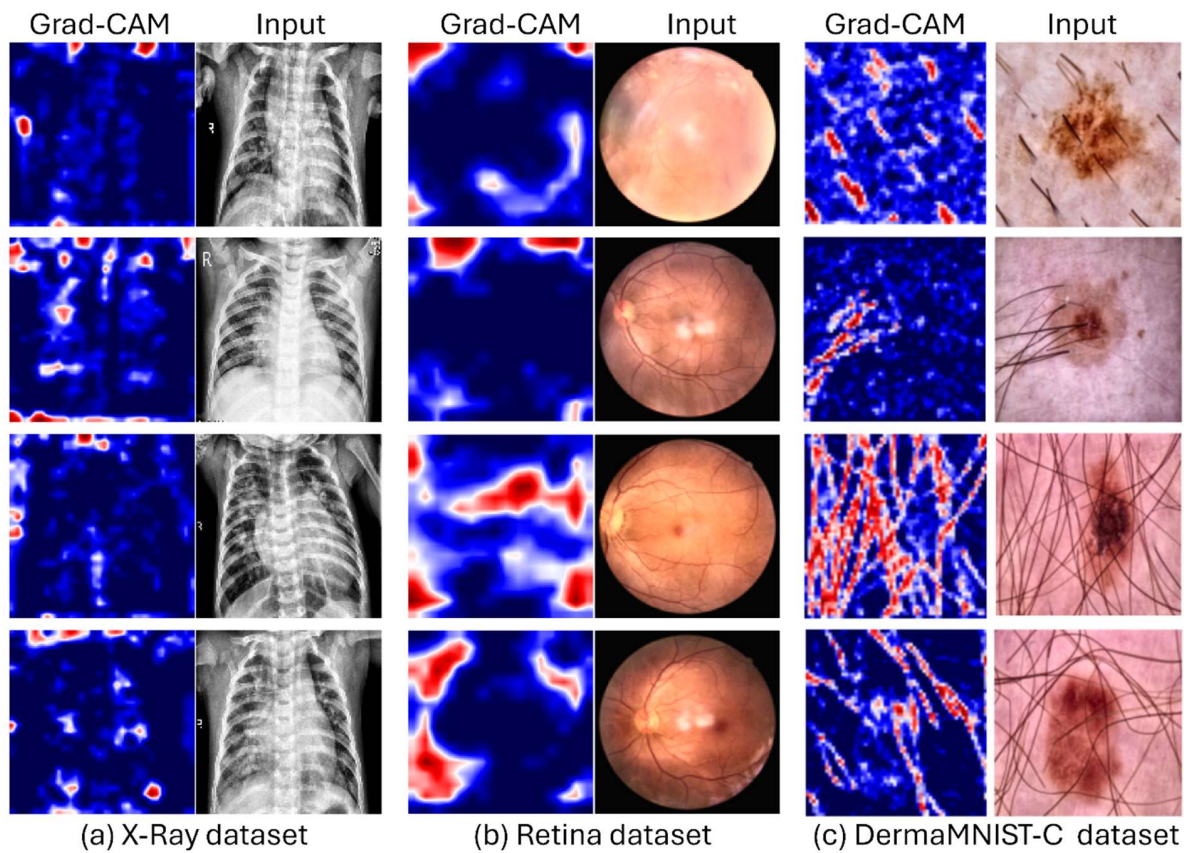

*Figure 1 Grad-CAM analysis on test images reveals bias in classification models. (left) Grad-CAM heatmap, which highlights activation. (right) Input image. This indicates the model's reliance on non-pathological cues. (A) Pneumonia X-ray [6]. (b) RetinaMNIST [7]. (c) Dermatological dataset DermaMNIST-C [8].*

To evaluate the impact of irrelevant features (see Figure 2and 2b), we applied a binary mask to remove the pixels corresponding to the "R" marker (see Figure 2c). This follows the principle of counterfactual explanations, in which minimal input modifications are used to test changes in model predictions. Once the marker was hidden, the model's confidence decreased significantly, indicating that it had learned to associate this non-diagnostic element with disease presence. Reliance on such spurious cues reflects dataset bias and introduces shortcuts that weaken the reliability of automated diagnostic systems.

In this study, we introduce the Guided Focus via Segment-Wise Relevance Network (GFSR-Net), a method that incorporates interpretability into the training of deep learning classifiers and evaluates its effectiveness on medical imaging tasks. GFSR guides the model's focus by a small number of human-provided annotations that approximate where attention should be directed within an image. This approach does not require precise boundaries or exhaustive coverage. These annotations generate non-binary relevance masks for all training images. We then incorporate these masks into a guided relevance loss that penalizes discrepancies between the target masks and the model's saliency maps computed via Grad-CAM. We validate our approach using chest X-rays, retinal scans, and dermatological images. The results demonstrate improved alignment between the model's focus and relevant regions, while maintaining comparable classification performance. A detailed description of the method and experimental setup can be found in sections 3 and 4.

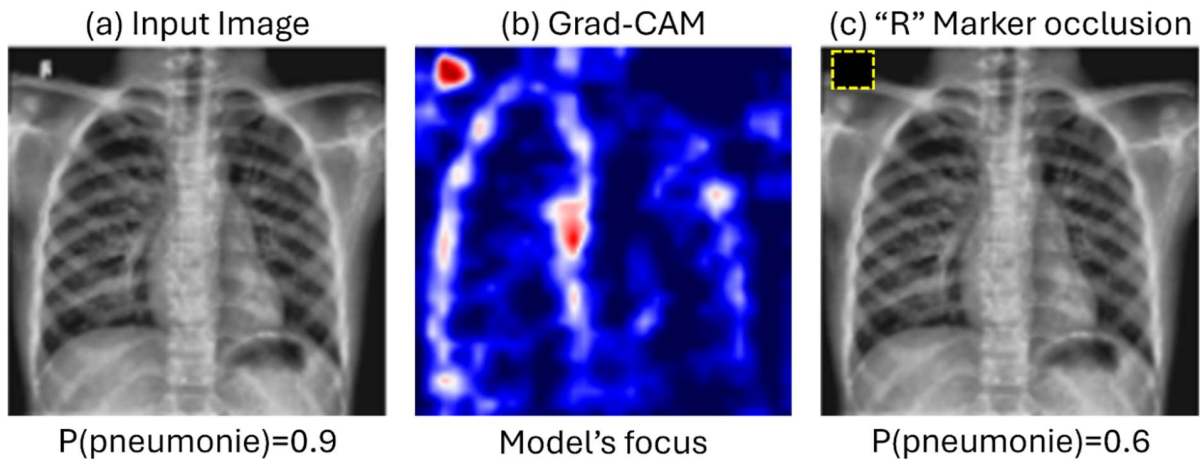

*Figure 2 Grad-CAM analysis reveals bias in a pneumonia classification model. (a) Input chest X-ray with a predicted probability of pneumonia of 0.9. (b) Grad-CAM heatmap, which highlights activation over an irrelevant region ("R" marker). (c) Occlusion of the "R" marker reduces the predicted probability to 0.6. This indicates the model's reliance on this non-pathological cue.*

## 2. Related work

In medical imaging, where transparency is essential for clinical adoption, applying interpretability methods to deep learning models has become a critical area of research. Post hoc XAI methods such as SHapley Additive exPlanations (SHAP) [9], Local Interpretable Model-Agnostic Explanations (LIME) [10], and Gradient-weighted Class Activation Mapping (Grad-CAM) [5], have been widely adopted to visualize the regions of an input image that influence a model's decision.

SHAP assigns an importance value to each feature based on Shapley values from cooperative game theory. It explains predictions by comparing the model's output with and without each feature across many combinations. LIME approximates the model locally using a simpler, more interpretable model (e.g., linear regression) on versions of the input that have been perturbed to explain individual predictions. Grad-CAM computes the gradient of the target class score with respect to feature maps in a convolutional layer. It then averages these gradients spatially to obtain importance weights, which it combines with the feature maps to generate a coarse heatmap. The resulting heatmap shows the regions that have the most significant influence on a model's prediction.

While these approaches focus on local explanations of individual predictions, another line of research explores concept-based methods for global interpretability. Concept-based interpretability methods are an emerging type of post-hoc XAI. Concept Activation Vectors (TCAV) [11] and Automated Concept-based Explanations (ACE) [12] are frameworks that associate human-understandable concepts with internal model representations and assess how sensitive a model's predictions are to specific concepts derived from the training data.

XAI methods allow clinicians to evaluate whether the model is focusing on the correct anatomical structures. However, these methods can only provide explanations after the model has been trained and do not influence the learning process.

Some studies have explored integrating attention or explanation mechanisms directly into the training process. Approaches such as the Convolutional Block Attention Module (CBAM)[13] and Attention Branch Networks (ABN) [14], [15] aim to refine feature extraction and improve classification and detection performance. However, these methods are usually developed and validated using large, unbiased datasets. These algorithms generate more localized, lower-entropy activation masks, which can mitigate some hidden biases, but may also inadvertently amplify others in small datasets. Other methods have proposed loss functions that align saliency maps with external expert-defined heatmaps or segmentations [16], [17]. However, these methods usually require a large number of well-defined annotations, which are costly and time-consuming to obtain.

In contrast, our method is designed for diagnostic imaging and does not rely on unsupervised mechanisms to generate attention. It also does not require large-scale or meticulously defined annotations on a per-image basis. Instead, we guide the network's focus using a limited set of images from which relevant segments are selected. This approach generates non-binary masks for each image in the training set, which are then incorporated into a mask-guided loss. Consequently, the model learns to focus on meaningful regions, striking a practical balance between interpretability, annotation effort, and robustness.

## 3. Methodology

In this section, we introduce the Guided Focus via Segment-Wise Relevance Network (GFSR-Net), a method designed to improve the interpretability and reliability of deep learning classifiers for medical imaging applications. Rather than relying solely on post-hoc explanations, GFSR-Net integrates human-guided annotations during training to direct the model's focus toward image regions that are intuitively relevant for diagnosis. The methodology is structured as follows: Section 3.1 outlines the GFSR-Net framework. Section 3.2 details the classification model architecture and experimental setup, and Section 3.3 introduces the evaluation datasets.

## 3.1 GFSR-Framework

The GFSR method combines segment-wise concept discovery with soft relevance supervision from a set of human annotations that approximate the natural visual focus a person would place on an image. These annotations offer coarse guidance by providing only positive relevance and do not require precise boundaries, exhaustive markings, or penalize unmarked regions. The overall framework is illustrated in Figure 3, showing how human annotations are translated into concept-level relevance masks that guide the model during training. The process involves segmenting each image into homogeneous regions using Simple Linear Iterative Clustering (SLIC) [18] and extracting feature embeddings with a ResNet50 backbone [19]. The resulting embeddings are then clustered into a fixed number of unsupervised concepts, each of which is assigned a score based on its overlap with the annotated regions. These scores generate non-binary relevance masks for the entire training set. Finally, a mask-guided relevance loss function penalizes discrepancies between the target masks and the saliency maps computed by Grad-CAM. This encourages the network to focus on regions that are diagnostically meaningful throughout the training process.

*(a)*

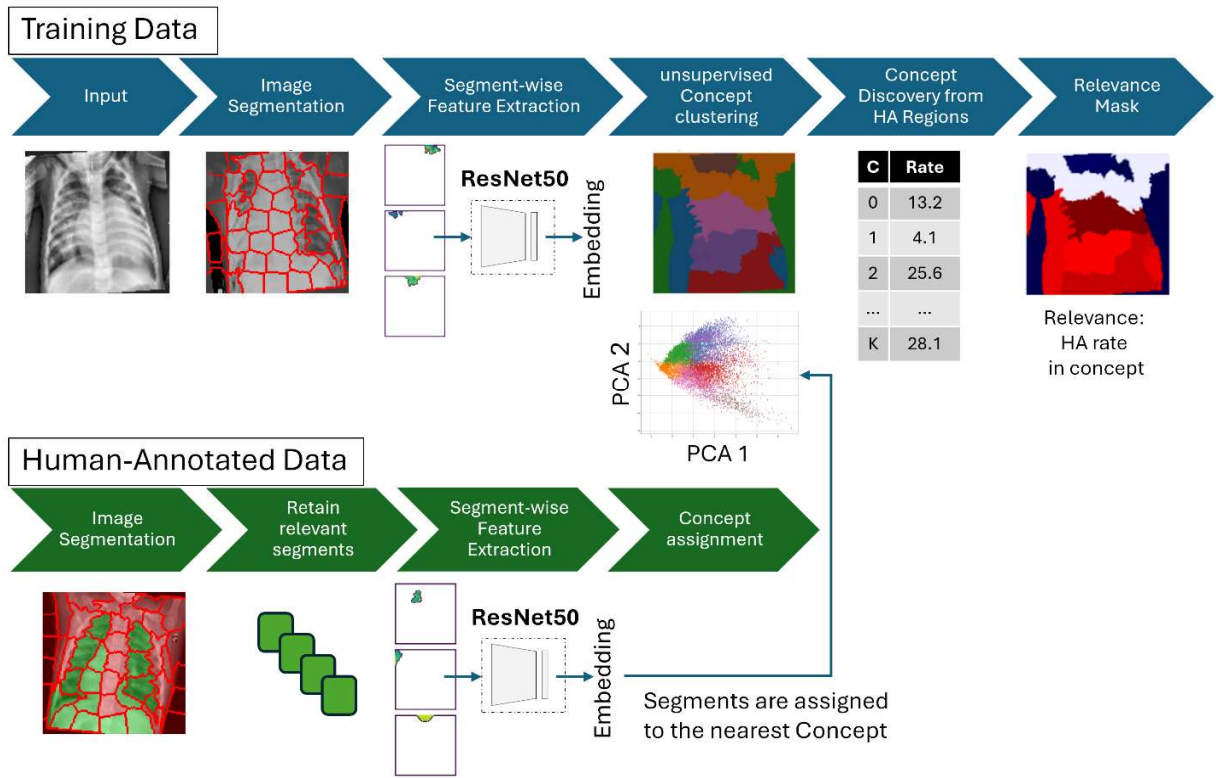

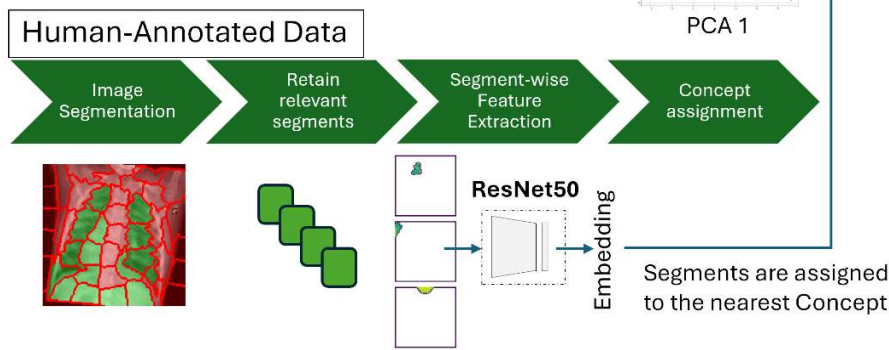

*(b)*

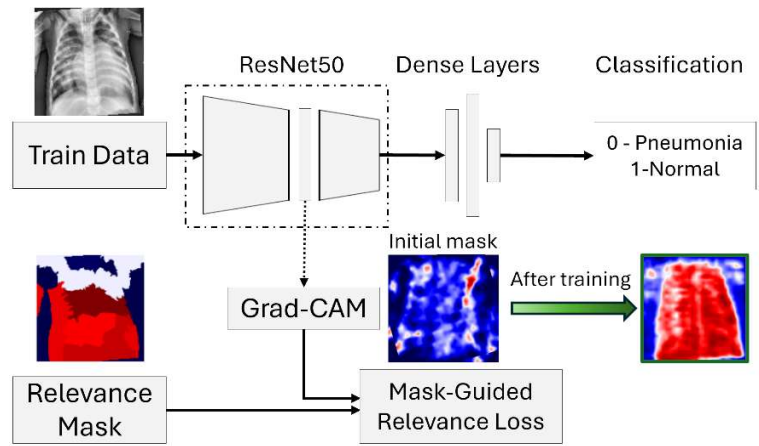

*Figure 3 Overview of the Proposed GFSR-Net Framework. (a) The framework uses two data sources: training images (top) and a small set of human-annotated (HA) images (bottom). The training images are segmented using the SLIC algorithm, and segment-wise embeddings are extracted using a ResNet50 backbone. These embeddings are then clustered into a fixed number of concepts. HA images provide soft supervision by highlighting relevant regions without requiring precise boundaries. Only the annotated segments are retained, and their embeddings define the relevance score of each concept. These scores are then propagated to all training segments via their concept assignment, generating non-binary relevance masks. (b) During training, the model's saliency maps (obtained with Grad-CAM) are aligned with the relevance masks via mask-guided relevance loss. This encourages attention to diagnostically meaningful regions. Unlike post hoc methods, saliency evolves dynamically during training under integrated, concept-level supervision.*

### 3.1.1 Image Segmentation and Feature Embedding Extraction

Each image is divided into visually coherent regions using the Simple Linear Iterative Clustering (SLIC) algorithm. This method produces compact, spatially connected superpixels that capture local image patterns. The compactness parameter controls the trade-off between color similarity and spatial proximity in SLIC segmentation. It was selected empirically based on visual inspection: 0.2 for chest X-rays, 10.0 for retinal images, and 25.0 for dermatological images. Each image was segmented into 50 regions to ensure consistent representation across datasets while adapting to the structural characteristics of each modality.

To represent features, we extract embeddings from each segment using a ResNet50 backbone pre-trained on ImageNet. The feature maps are max-pooled to produce a fixed-length vector for each segment, capturing high-level visual characteristics for subsequent concept discovery. For human-annotated images, only the relevant segments are retained for further processing. Figure 3a illustrates the segmentation and embedding extraction steps.

### 3.1.2 Concept Discovery from Human-Annotated Regions

To guide the model toward meaningful areas of an image, we use a small set of human-annotated images that provide soft relevance supervision. These annotations highlight regions that are visually relevant from a human perspective without requiring precise boundaries or full coverage. Segments that overlap with the annotated regions are considered relevant, while areas that are not annotated are ignored, even if they contain additional information

As shown in Figure 3a, the training images and the HA images are segmented using SLIC. Then, segment embeddings are extracted as described in Section 3.1.1. For HA images, only the embeddings that correspond to the annotated segments are retained. These embeddings are then clustered into a fixed number of unsupervised concepts using k-means clustering (we use seven clusters), similar to TCAV [11] or ACE [12]. The resulting concepts do not carry predefined semantic meaning, but instead group visually similar regions based on embedding similarity.

Each concept receives a relevance score proportional to the fraction of annotated segments it contains from HA images. These scores are propagated to all segments in the training set according to their concept membership. This results in non-binary relevance masks for each image. These masks indicate which regions should be prioritized during training while allowing flexibility for unmarked areas.

This strategy enables concept-level supervision with only a few annotated examples. In our experiments, annotating as few as five images per class was sufficient due to low intra-class variability. For instance, in pneumonia classification, the lung area consistently contains the most informative features, rendering sparse, approximate guidance highly effective. Furthermore, perfect masks are unnecessary because mask-guided relevance loss (described in Section 3.1.3 and illustrated in Figure 3b) tolerates imprecise boundaries and operates on averaged differences between target and predicted relevance.

### 3.1.3 Mask-Guided Relevance Loss

The mask-guided relevance loss compares the model's saliency maps, which are computed via Grad-CAM during training, with non-binary relevance masks that are derived from concept-level scores (see Figure 3b). This alignment process strengthens activation in regions associated with human-provided annotations and suppresses focus on irrelevant or misleading areas.

Specifically, the Grad-CAM maps are extracted from the third convolutional block of the ResNet50 backbone. This provides a spatial estimate of the regions that are most influential to the model's predictions at an intermediate feature level. Each training image has an associated soft relevance mask constructed by propagating concept relevance scores to its segments according to their cluster assignments.

The relevance loss is formulated as a weighted combination of three terms.

$$\mathcal{L}_{relevance} = \alpha \cdot MAE + (1 - \alpha) \cdot MSE + \beta \cdot Focal$$

where:

- The mean absolute error (MAE) measures the average absolute deviation between the attention map and the expected mask.

- The mean squared error (MSE) penalizes large deviations more heavily.

- Focal increases the penalty in areas with high expected relevance, ensuring that omissions in critical regions (e.g., lungs) are more heavily weighted.

- $\alpha$ controls the balance between MAE and MSE, and $\beta$ determines the strength of the focal term.

The total loss function used during training combines the relevance loss with the standard classification loss:

$$\mathcal{L}_{total} = \mathcal{L}_{cls} + \mathcal{L}_{relevance}$$

where $\mathcal{L}_{cls}$ corresponds to binary cross-entropy in binary classification tasks, or sparse categorical cross-entropy in multi-class settings.

## 3.2 Classification Model Architecture and Setup

We use a pretrained ResNet50 backbone on ImageNet [19], a well-established architecture that requires minimal parameter tuning. As described in Section 3.1, the backbone supports the extraction of segment-wise embeddings and integrates with the proposed mask-guided relevance loss. Our primary objective is to enhance interpretability while maintaining competitive predictive performance, rather than optimizing solely for classification accuracy.

To obtain a compact latent representation, we apply a GlobalMaxPool2D layer to the feature maps produced by the backbone, obtaining a fixed-length embedding vector. This vector is then passed through a dense layer containing 128 neurons, followed by a dropout layer with a rate of 0.3 to reduce overfitting. A fully connected output layer produces the class predictions.

To isolate the effect of the mask-guided relevance loss on interpretability, we use the same model architecture for all datasets and compare its performance with an identical model that was trained without relevance loss. Notably, the baseline models converge much faster, requiring only about 100 to 500 epochs, whereas our approach needs approximately 500 to 1000. Despite this difference in training efficiency, the inference time remains unchanged.

## 3.3 Datasets

We evaluated our method on three publicly available medical imaging datasets: Pneumonia X-ray [6], RetinaMNIST [7] from the MedMNIST v2 collection, and the corrected dermatological dataset DermaMNIST-C [8]. Pneumonia X-ray and RetinaMNIST images are resized to 128x128 pixels, while DermaMNIST-C images are resized to 224x224 pixels. Preprocessing follows the standard Keras ResNet50 pipeline, converting images from RGB to BGR and subtracting ImageNet mean values (103.939 for blue, 116.779 for green, and 123.68 for red). For grayscale Pneumonia X-ray images, the single channel is replicated across three channels to meet ResNet50's input requirements. RetinaMNIST and DermaMNIST-C images are already in RGB format and require no adjustment.

The pneumonia dataset contains 5,216 training images and 624 test images labeled as either normal or pneumonia. The RetinaMNIST dataset includes 1,080 training images and 400 test images labeled across five categories of diabetic retinopathy severity: no DR, mild, moderate, severe, and proliferative. DermaMNIST-C consists of 8,215 training images and 1,227 test images across seven diagnostic categories. In the test set, the class distribution is as follows: actinic keratoses and intraepithelial carcinoma (35), basal cell carcinoma (44), benign keratosis-like lesions (105), dermatofibroma (8), melanoma (70), melanocytic nevi (948), and vascular lesions (17). The dataset is highly imbalanced, with melanocytic nevi dominating both training and test splits. We use the predefined training and test splits provided with each dataset to ensure consistency with prior work and enable fair comparisons.

Although we report standard classification metrics to verify baseline performance, our main focus is evaluating whether the model aligns its internal representations with human-guided annotations through concept-level supervision and mask-guided relevance loss. In addition to these benchmarks, In Section 4.1, we provide a controlled example to demonstrate the effect of non-causal correlations on our method.

## 4. Results and Discussion

This section begins with a controlled example that illustrates the principle of our method (Section 4.1). This is followed by a quantitative evaluation of our method on medical imaging datasets (Section 4.2). Finally, we analyze the interpretability and robustness of our model compared to a baseline model (Section 4.3).

### 4.1 Wolf vs. Husky Background Bias dataset

As described in Section 3, our method guides the model's focus toward relevant regions rather than irrelevant cues. To demonstrate this principle in a controlled setting, we adapted the well-known Husky vs. Wolf experiment [10]. The training set consists of 11 wolf images in snowy environments and 11 Husky images against non-snowy backgrounds. This creates a background bias that often causes standard neural networks to misclassify snow as part of the "wolf" category.

During training, GFSR-Net was guided using a few manually selected segments corresponding to the animals' bodies ( Figure 4b). These annotations were used to generate non-binary relevance masks that directed the model's attention.

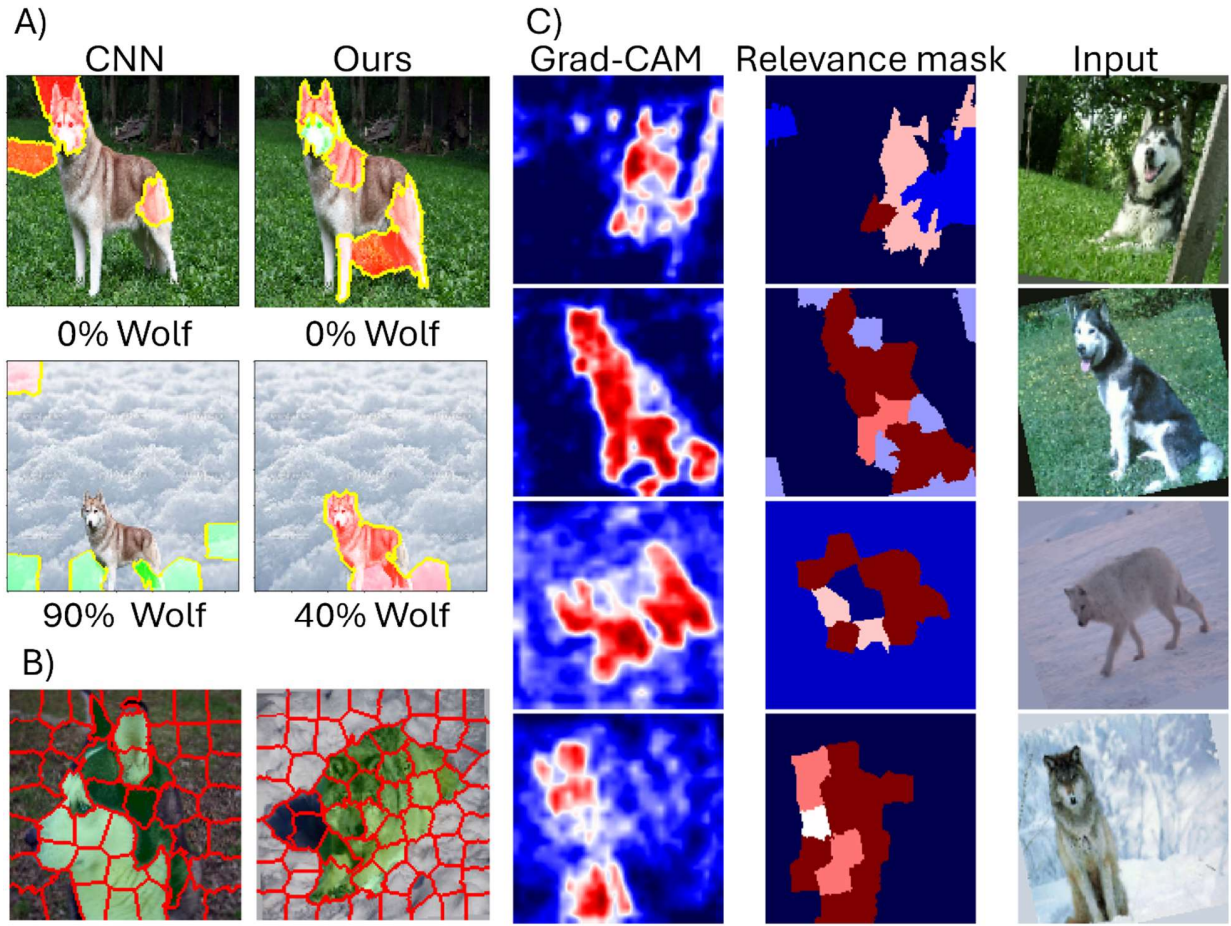

*Figure 4 (a) LIME compares a standard CNN (left) and our method (right) on the dataset of wolf versus Husky. Green and red highlights indicate regions supporting the "Wolf" and "Husky" classes, respectively. The top row shows a Husky on a neutral background, correctly classified as 0% wolf by both models. The bottom row shows the same dog on a snowy background. Standard CNN relies heavily on background cues and classifies the dog as 90% wolf, whereas our method focuses on the animal's body and reduces the bias, classifying the dog as 40% wolf. (b) Manually selected "useful segments". (c) Comparison of Grad-CAM saliency maps (left), segment-based focus masks (middle), and the original input images (right) for different examples. The focus masks highlight the regions that were emphasized during training, and the Grad-CAM shows the areas of activation that contributed to the model's prediction.*

We further assessed this behavior with LIME (Figure 4a). In a critical test case, the baseline model misclassified a dog in a snowy environment as a "wolf" (Figure 4a, bottom left) due to its reliance on the background. In contrast, GFSR-Net correctly predicted "Husky" (Figure 4a, bottom right) based on body features highlighted in the relevance masks. Figure 4c compares Grad-CAM saliency maps (left), relevance masks (middle), and input images (right). While baseline Grad-CAM maps highlight snow-dominated backgrounds, our mask-guided relevance loss shifts attention to the animal's body, avoiding reliance on irrelevant context.

Although background bias could be mitigated with preprocessing (e.g., foreground masking), we intentionally retained the original images. This demonstrates that our mask-guided relevance supervision alone can prevent misleading correlations, maintaining a simple and accessible approach without the need for additional preprocessing steps.

## 4.2 Classification Performance

We evaluated the classification performance of GFSR-Net using the Pneumonia X-ray, RetinaMNIST, and DermaMNIST-C datasets. Table 1, Table 2, and Table 3 summarize the results in terms of the area under the curve (AUC), as well as accuracy, sensitivity, and specificity. To ensure a fair comparison, we used the same backbone and dense layers as the ResNet50 baseline model; the only difference was the inclusion of mask-guided relevance loss in our approach. We took performance values for additional methods from the literature, assuming the same predefined training, validation, and test splits.

For pneumonia classification, GFSR-Net achieved an AUC of 0.961, whereas the baseline achieved an AUC of 0.958 (Figure 5a). Though the numerical difference is slight, GFSR-Net outperformed the baseline consistently across most false positive rates, achieving a better balance between sensitivity and specificity. Table 1 shows that the baseline achieved perfect sensitivity (1.0) at the expense of low specificity (0.491), whereas GFSR-Net provided a more balanced trade-off (sensitivity 0.979 and specificity 0.730).

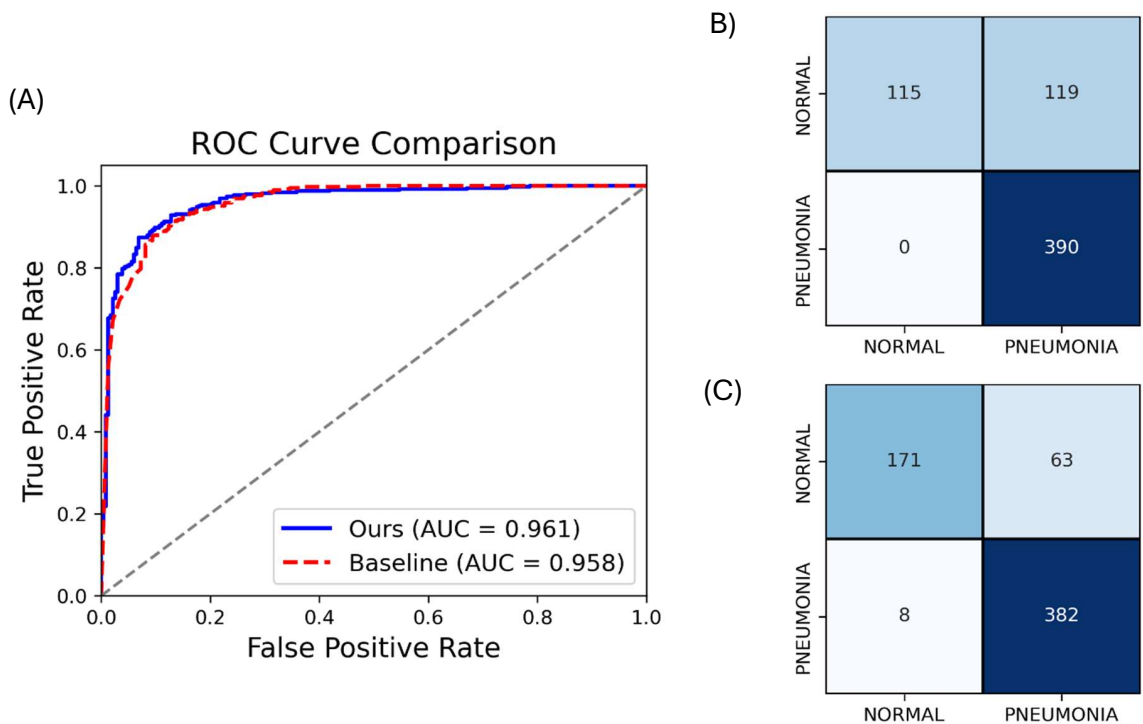

*Figure 5 (A) Receiver operating characteristic (ROC) curves comparing the proposed and baseline models for pneumonia classification. The proposed model achieved an area under the curve (AUC) of 0.961, slightly outperforming the baseline model (AUC = 0.958) with consistently higher true positive rates across most false positive rates. (B) Confusion matrix of the baseline model. (C) Confusion matrix of the proposed model.*

Table 1 *Performance comparison between the baseline ResNet-50 model and the proposed Guided Focus Network (GF-Net) in terms of sensitivity, specificity, and overall accuracy and AUC for the pneumonia dataset.*

| Model | Sensitivity | Specificity | Accuracy | AUC |
|---|---|---|---|---|
| ResNet-50 (base) | **1.000** | 0.491 | 0.809 | 0.958 |
| ResNet-50 (GF-Net) | 0.979 | **0.730** | **0.886** | **0.961** |

On RetinaMNIST, GFSR-Net achieved an area under the curve (AUC) of 0.797 and an accuracy of 0.578, compared to the ResNet50 baseline with an AUC of 0.817 and an accuracy of 0.603 (see Table 2). While these metrics suggest a slight decrease overall, this is largely due to the significant class imbalance in the test set, in which the dominant class comprises 174 out of 400 samples. Other measures indicate that GFSR-Net outperforms the baseline: sensitivity increased from 0.46 to 0.50, precision remained at 0.50, and the F1 score improved from 0.47 to 0.49. As shown in Figure 6, the baseline model tends to concentrate predictions on the majority class, resulting in poor recognition of mild and moderate diabetic retinopathy. In contrast, GFSR-Net distributes attention more evenly across classes. This improves sensitivity to early disease while maintaining solid performance in the no DR and severe categories.

Table 2 *Performance comparison between the baseline ResNet-50 model and the proposed Guided Focus Network (GF-Net) in terms of sensitivity, specificity, and overall accuracy (ACC) for the RetinaMNIST dataset.*

| Model | Sensitivity | Precision | F1-score | Accuracy | AUC |
|---|---|---|---|---|---|
| ResNet-50 (base) | 0.46 | 0.50 | 0.47 | **0.603** | **0.817** |
| ResNet-50 (GF-Net) | **0.50** | 0.50 | **0.49** | 0.578 | 0.797 |

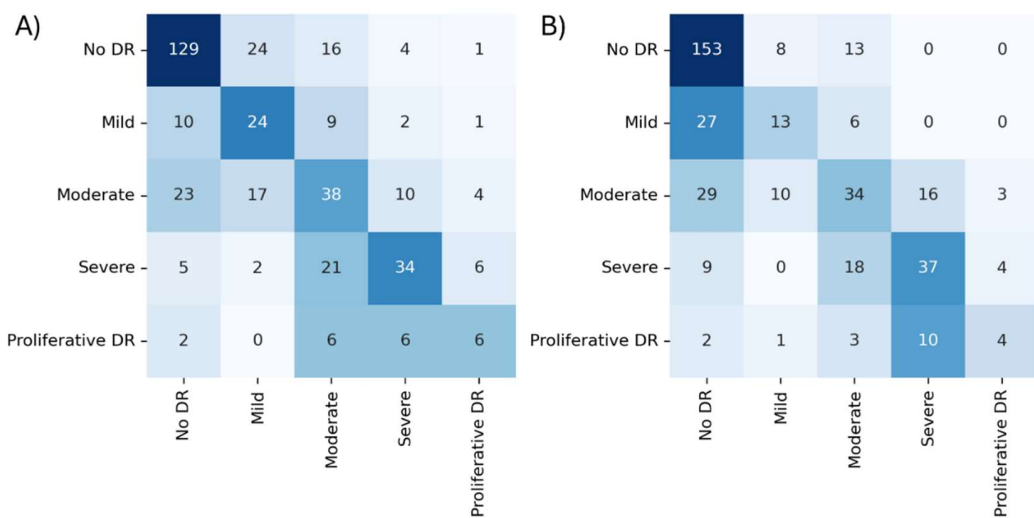

Figure 6 *Confusion matrices on the diabetic retinopathy dataset. (A) proposed method (GF-Net). (B) baseline method. The proposed method achieves higher sensitivity, precision, and F1-score. This shows improved recognition across all disease grades, particularly for mild and moderate cases, while maintaining strong performance in the no DR and severe classes.*

On the DermaMNIST-C dataset, GFSR-Net achieved an area under the curve (AUC) of 0.936 and an accuracy of 0.869. In comparison, the baseline ResNet-50 achieved an AUC of 0.947 and an accuracy of 0.870 (see Table 3). Although the overall metrics are nearly identical, GFSR-Net shows consistent improvement in sensitivity (0.57 to 0.59), precision (0.67 to 0.70), and F1 score (0.61 to 0.62). Confusion matrices in Figure 7 further highlight these differences: the baseline model is dominated by majority class predictions (melanocytic nevi), reducing its ability to detect less represented categories. In contrast, GFSR-Net more clearly recognizes minority classes, such as actinic keratoses, intraepithelial carcinoma, basal cell carcinoma, and vascular lesions. These results suggest that mask-guided relevance supervision improves category balance, mitigating the bias toward the most frequent class and providing more reliable predictions for rare but clinically significant cases.

*Table 3 Performance comparison between the baseline ResNet-50 model and the proposed Guided Focus Network (GF-Net) in terms of Sensitivity, precision, F1-score, overall accuracy and AUC for the DermaMNIST dataset-C*

| Model | Sensitivity | Precision | F1-score | Accuracy | AUC |
|---|---|---|---|---|---|
| ResNet-50 (base) | 0.57 | 0.67 | 0.61 | **0.870** | **0.947** |
| ResNet-50 (GF-Net) | **0.59** | **0.70** | **0.62** | 0.869 | 0.936 |

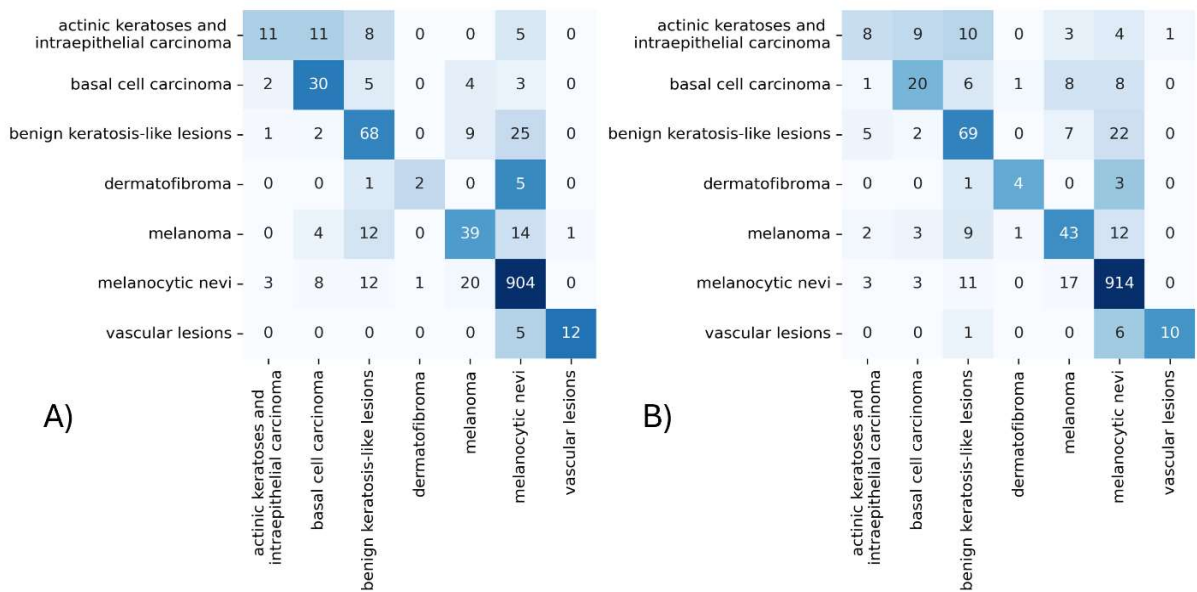

*Figure 7 Confusion matrices on the diabetic DermaMNIST dataset-C. (A) proposed method (GF-Net). (B) baseline method. The proposed method achieves higher sensitivity, precision, and F1-score. This shows improved recognition across some classes, particularly for actinic keratoses and intraepithelial carcinoma, basal cell carcinoma, and vascular lesions.*

Previous results confirm that incorporating soft relevance supervision preserves discriminative performance and can offer measurable improvements, especially in complex, imbalanced, or multi-class classification tasks.

## 4.3 Interpretability and Robustness

### 4.3.1 Analysis of Individual Images

We first analyzed individual images to evaluate GFSR-Net attention compared to the baseline. As illustrated in Figure 8, the input images were segmented into regions, clustered into unsupervised concepts, and assigned relevance scores based on their overlap with the human-provided annotations. These scores generate non-binary relevance masks that guide the network during training. The resulting Grad-CAM saliency maps confirm that GFSR-Net focuses on anatomically meaningful structures. This demonstrates that concept-guided supervision effectively directs the model toward relevant regions while suppressing irrelevant ones.

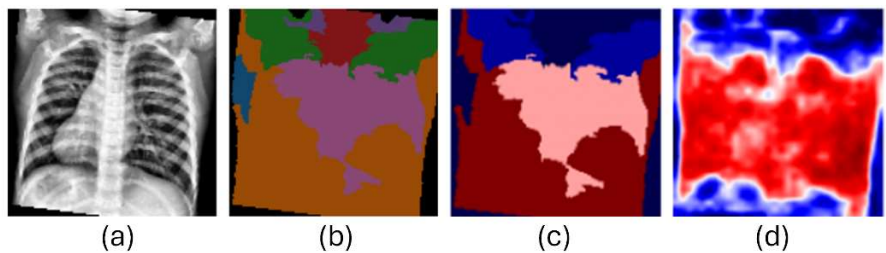

*Figure 8 (a) original input X-ray image. (b) segmented image showing unsupervised concept regions. (c) concept-guided relevance mask used for supervision. (d) Grad-CAM saliency map highlighting the model's focus.*

A direct comparison of Grad-CAM visualizations further highlights these differences (Figure 9). Each row contains three columns. The first column shows the Grad-CAM visualization generated by GFSR-Net. The second column shows the Grad-CAM of the baseline model. The third column contains the input image. As can be seen, the interpretations produced by GFSR-Net are more consistent and anatomically meaningful, whereas the baseline Grad-CAM maps are more diffuse and difficult to interpret.

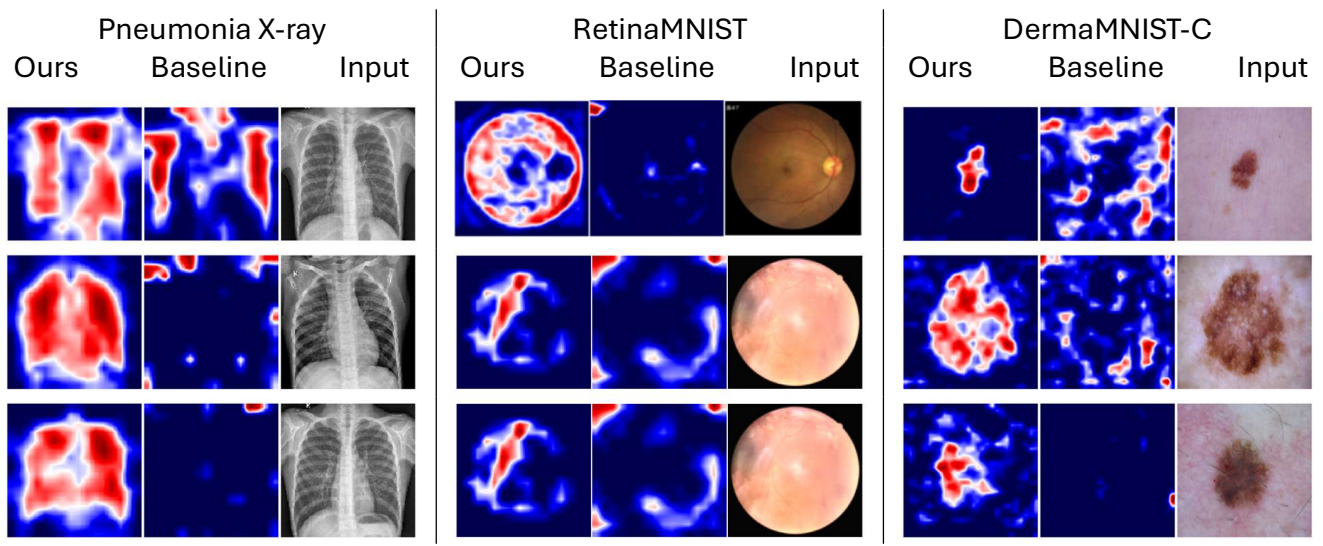

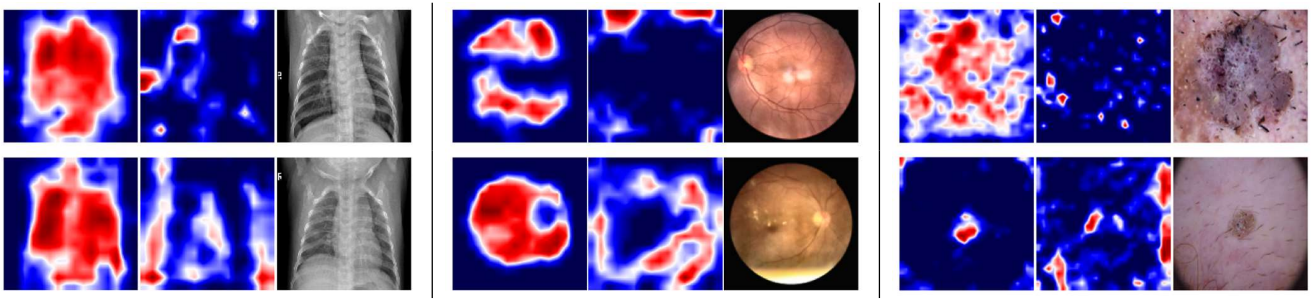

Figure 9 Comparison of Grad-CAM visualizations between GFSR-Net and the baseline model. Each row shows the Grad-CAM generated by GFSR-Net on the left, the Grad-CAM generated by the baseline model in the middle, and the corresponding input image on the right. GFSR-Net produces clearer, more anatomically meaningful attention maps. In contrast, the baseline model focuses on less relevant or more diffuse regions, making its interpretation more challenging. (a)

In chest X-rays (Figure 9a), baseline activations often concentrated on irrelevant artifacts such as the 'R' marker or text annotations, whereas GFSR-Net consistently emphasized pulmonary regions. We examined the model's robustness under spatial modifications of chest X-rays (Figure 10). Grad-CAM visualizations of the baseline model revealed attention to irrelevant artifacts, such as the "R" marker in the upper-left corner and textual annotations in the lower-right corner. Cropping these images caused large fluctuations in the baseline model's predictions, indicating its sensitivity to spurious cues. For instance, the presence of the "R" marker increased the likelihood of a pneumonia diagnosis, whereas the lower-right annotation favored a normal prediction. In contrast, GFSR-Net maintained stable predictions by focusing on pulmonary regions.

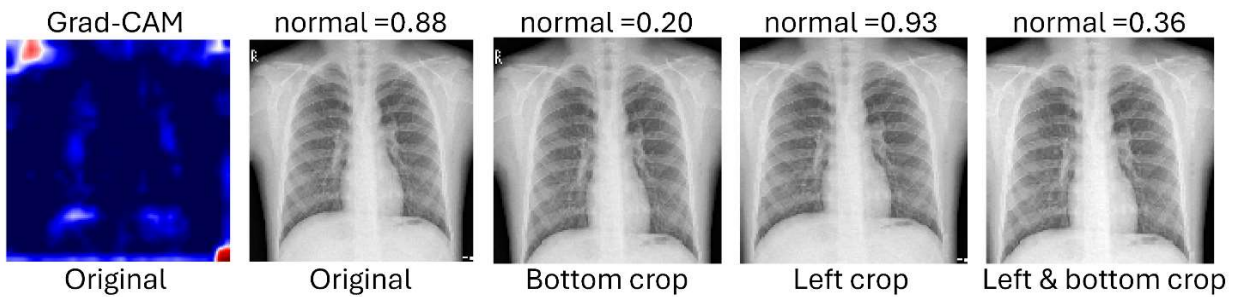

Figure 10 Grad-CAM visualization and based-model predictions under different cropping conditions. The first image displays the Grad-CAM heatmap for the original chest X-ray, highlighting the most influential regions contributing to the prediction. Subsequent images show the same X-ray with different crops (original, bottom, left, and left- bottom) and the corresponding predicted probability of the "normal" class. Changes in prediction values indicate the effect of spatial context on the model's confidence.

In the case of retinal images (Figure 9b), the baseline model sometimes relied on irrelevant features, such as corner annotations or the optic disc, which is an unrelated structure. Its activations were often sparse and scattered across the background. In contrast, GFSR-Net consistently highlighted broader retinal areas, demonstrating reduced reliance on misleading cues. An occlusion test confirmed this behavior: Masking a text annotation shifted the baseline prediction from "mild DR" to "no DR," whereas GFSR-Net's prediction remained unchanged.

For dermatological images (Figure 9c), the baseline model frequently emphasizes irrelevant areas, such as the surrounding skin, background textures, and dark hairs, instead of the lesion itself. These diffuse and inconsistent activations reduced the interpretability of the predictions and increased the risk of bias from spurious features. In contrast, GFSR-Net produced clearer, more localized saliency maps that aligned with the lesion area, reducing reliance on peripheral artifacts.

### 4.3.2 Robustness under Perturbation

In addition to individual examples, we next assessed robustness under controlled perturbations. Robustness was quantified using two complementary metrics: the agreement rate, indicating consistency between original and perturbed predictions, and the flip rate, capturing prediction instability. Starting with chest X-rays, we applied cropping operations that preserved the clinical content while removing annotations that could otherwise bias the predictions. On this dataset, the baseline model achieved an agreement rate of 93.9% (with a flip rate of 6.1%). The confusion matrix of flips revealed marked asymmetry, with 27 predictions shifting from normal to pneumonia and 11 shifting in the opposite direction. In contrast, GFSR-Net achieved a higher agreement rate of 97.6% (with a flip rate of 2.4%) and had far fewer flips overall (seven from normal to pneumonia and eight from pneumonia to normal).

For retinal images, cropping was not a suitable option to remove potential annotations, as it risked eliminating diagnostically relevant regions. Instead, we used a selective blurring strategy to evaluate robustness while preserving the central retinal structures. Specifically, we automatically identified the center of the eyeball and applied a Gaussian blur with a standard deviation of 1.5 and a feather of 2 outside a radius of 60 pixels. This procedure preserved clinically meaningful information in the central retina while gradually obscuring peripheral artifacts and annotations. With this approach, GFSR-Net achieved an agreement rate of 0.853 and a flip rate of 0.148 compared to baseline rates of 0.823 and 0.178, respectively.

For DermaMNIST-C, we applied a similar perturbation strategy (see Figure 11). Instead of Gaussian blurring, we replaced the peripheral pixels with the background of validation image 25, which shows a benign keratosis-like lesion. The lesion itself is not visible in this reference image, but several dark hairs are present. As illustrated in Figure 1c, these hairs introduce bias into the dataset. With this setup, GFSR-Net achieved an agreement rate of 0.667 and a flip rate of 0.333. In contrast, the baseline achieved a much lower agreement rate of 0.315 and a considerably higher flip rate of 0.685. The confusion matrices of flips further emphasize this contrast; the baseline model shows large, systematic shifts dominated by hair-related artifacts, whereas GFSR-Net produces more balanced predictions across categories. These results suggest that relevance-guided training enhances robustness by mitigating the impact of irrelevant features, such as hair.

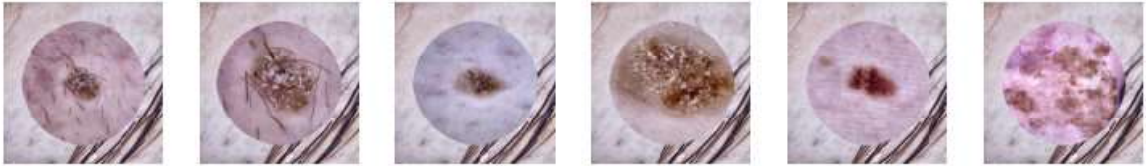

*Figure 11 Examples of dermatological images from DermaMNIST-C containing visible hair artifacts. These images were used to test the influence of spurious features, such as dark hairs, which can introduce bias into the classification. In our robustness experiments, peripheral regions containing these artifacts were replaced with background patches to evaluate the stability of baseline and GFSR-Net predictions.*

## 5. Conclusions

We introduced GFSR-Net, a framework that integrates focus guidance into the training process of deep learning models, improving interpretability and robustness, especially in sensitive domains such as medical imaging. The model is encouraged to attend to meaningful regions (e.g., anatomical structures) rather than irrelevant artifacts by incorporating non-binary relevance masks.

Controlled experiments with natural images showed that GFSR-Net can overcome misleading correlations, such as a snow-dominated background in the Husky vs. Wolf example, by focusing on the animal itself. Results on medical datasets confirmed that the method preserves strong predictive performance while producing more interpretable and stable attention patterns.

A key advantage of the approach is its efficiency; only a small number of coarse annotations are needed. Through concept clustering, relevance information propagates across the dataset, automatically generating soft supervision that is scalable and noise tolerant.

Unlike traditional post hoc explainability methods, GFSR-Net dynamically shapes saliency during training via mask-guided relevance loss, aligning the model's internal focus with human-guided concepts. This provides relevance maps alongside predictions, strengthening transparency and trustworthiness without requiring architectural changes or sacrificing accuracy. Future work may extend this strategy to other imaging modalities and tasks and integrate it with uncertainty quantification to provide even more reliable clinical decision support.

Robustness experiments across Pneumonia, RetinaMNIST, and DermaMNIST-C demonstrate that GFSR-Net aligns attention with human-annotated regions. This reduces reliance on spurious cues, such as annotations, image borders, and hair artifacts. By directing the model toward areas that humans intuitively deem relevant, this approach mitigates prediction instability under perturbations, preserves diagnostic content, and enhances recognition of minority classes. In this study, the annotations were provided by non-expert users rather than medical experts, which resulted in measurable gains. Therefore, it can be expected that using expert annotations and larger annotated datasets would further enhance the approach's effectiveness. These results demonstrate that non-binary relevance masking improves interpretability and robustness against dataset-specific biases, thereby increasing the reliability of automated medical image analysis.